\documentclass[twocolumn,secnumarabic,amssymb, nobibnotes, aps, prd]{revtex4}
\usepackage{graphicx}
\usepackage{amsmath,dsfont}
\usepackage{overpic}

\newcommand{\beq}{\begin{equation}}
\newcommand{\eeq}{\end{equation}}
\newcommand{\beqa}{\begin{eqnarray}}
\newcommand{\eeqa}{\end{eqnarray}}
\newcommand{\no}{\nonumber}

\newcommand{\sfrac}[2]{{\textstyle\frac{#1}{#2}}}

\newcommand{\mnod}{M_0} 

\begin{document}
\title{\Large\bf Kaonic hydrogen and $\mbox{\boldmath$K^- p$}$ scattering }
\vskip 1.5true cm
\author{B. Borasoy$^{a,b}$}
\email{borasoy@itkp.uni-bonn.de}
\author{R. Ni{\ss}ler$^{a,b}$}
\email{rnissler@itkp.uni-bonn.de}
\author{W. Weise$^{a}$}
\email{weise@ph.tum.de}
\affiliation{\it $^{a}$ Physik Department, Technische 
Universit\"at M\"unchen, D-85747 Garching, Germany}
\affiliation{\it $^{b}$ Helmholtz-Institut f\"ur Strahlen- und Kernphysik (Theorie),
             Universit\"at Bonn, Nu{\ss}allee 14-16, D-53115 Bonn, Germany}

\begin{abstract}
Chiral SU(3) effective field theory in combination with a relativistic coupled channels 
approach is used to perform a novel analysis of the strong interaction shift and width 
in kaonic hydrogen in view of the new accurate DEAR measurements. Questions of consistency 
with previous $K^- p$ data are examined. Coulomb and isospin breaking effects turn out to 
be important and are both taken into account in this work.

\begin{center}
\end{center}
\end{abstract}

% 11.80.-m   Relativistic scattering theory 
% 11.80.Gw   Multichannel scattering
% 12.39.Fe   Chiral Lagrangians
% 13.75.Gx   Pion-baryon interactions
% 13.75.Jz   Kaon-baryon interactions
% 36.10.Gv   Mesonic atoms and molecules, hyperonic atoms and molecules

\maketitle

%%%%%%%%%%%%%%%%%%%%%%%%%%%%%%%%%%%%%%%%%%%%%%%%%%%%%%%%%%%%%%%%%%%%%%%%%%%%%%

The low-energy $\bar{K}N$ system is of special interest as a testing ground for 
chiral SU(3) symmetry in QCD and, in particular, for the role of explicit symmetry 
breaking induced by the relatively large mass of the strange quark. 
Most significantly, the existence of the $\Lambda(1405)$ resonance just  25 MeV 
below the $K^- p$ threshold makes chiral perturbation theory inapplicable in this channel. 
Non-perturbative coupled-channel techniques based on driving terms of the chiral SU(3) 
effective Lagrangian have proved useful and successful in dealing with this problem, 
by generating the $\Lambda(1405)$ dynamically as an I = 0  $\bar{K}N$ quasibound state 
and as a resonance in the $\pi\Sigma$ channel. High-precision $K^- p$ threshold data set 
important constraints for such theoretical approaches. Now that new accurate results for 
the strong interaction shift and width of kaonic hydrogen from the DEAR experiment 
\cite{DEAR} are available, there is renewed interest in an improved analysis of these 
data together with existing information on $K^- p$ scattering, the $\pi\Sigma$ mass 
spectrum and $K^- p$ threshold decay ratios.

The combination of chiral SU(3) effective field theory with coupled channels was first introduced in 
ref.~\cite{KSW1} and subsequently further developed and applied to a variety of meson-baryon scattering 
and photoproduction processes \cite{KSW2, OR, OM, LK, BMW, JOORM}. 
The starting point of this coupled channels approach is the chiral effective
Lagrangian which incorporates the same symmetries and symmetry breaking patterns as
QCD and describes the coupling of
the pseudoscalar meson octet $(\pi,K,\eta)$ to the ground state
baryon octet $(N,\Lambda, \Sigma, \Xi)$: 
\begin{equation}
{\cal L} = {\cal L}_\phi + {\cal L}_{\phi B}\,.
\end{equation}

The purely mesonic part of the Lagrangian ${\cal L}$ up to second chiral order
is given by ${\cal L}_\phi$ \cite{GL}, while 
the second part ${\cal L}_{\phi B}$ describes the meson-baryon
interactions and reads at lowest order \cite{K}
\begin{eqnarray}  \label{bar}
{\cal L}_{\phi B}^{(1)} &=& i \langle \bar{B} \gamma_{\mu} [D^{\mu},B] \rangle 
 - \mnod \langle \bar{B}B \rangle \no \\
&-&   \frac{1}{2} D \langle \bar{B} \gamma_{\mu}
 \gamma_5 \{u^{\mu},B\} \rangle  
- \frac{1}{2} F \langle \bar{B} \gamma_{\mu} \gamma_5 [u^{\mu},B] \rangle 
\end{eqnarray}
with $\langle \ldots \rangle$ denoting the trace in flavor space.
The pseudoscalar meson octet $\phi$ is summarized in
$u_{\mu} = i u^\dagger \partial_{\mu} U u^\dagger$
where $U = u^2 = \exp (\sqrt{2} i\phi / f )$,
and $f$ is the pseudoscalar decay constant in the chiral limit.
The ground state baryon octet is collected in the 
$3 \times 3$ matrix $B$, $\mnod$ is the common baryon octet mass in the
chiral limit and $D,F$ are the axial vector couplings of the baryons
to the mesons. The values of $D$ and $F$ are extracted from the empirical 
semileptonic hyperon decays. A fit to data gives $D= 0.80 
\pm 0.01$, $F=0.46 \pm 0.01$ \cite{CR}. 
Finally, the covariant derivative of the baryon fields is
\begin{equation}
[ D_\mu, B] = \partial_\mu B + [ \Gamma_\mu, B] 
\end{equation}
with the chiral connection
\begin{equation} \label{gama}
\Gamma_\mu = \sfrac{1}{2} [ u^\dagger,  \partial_\mu u]  .
\end{equation}

At next-to-leading order the terms relevant for $s$-wave meson-baryon
scattering are
\begin{eqnarray}  \label{bar2}
{\cal L}_{\phi B}^{(2)} &=&  b_D \langle \bar{B}  \{\chi_+,B\} \rangle +
b_F \langle \bar{B}  [\chi_+,B] \rangle + b_0 \langle \bar{B}B \rangle
\langle \chi_+ \rangle \no \\
&&
+ d_1 \langle \bar{B} \{u_{\mu},[u^{\mu},B]\}  \rangle 
+ d_2 \langle \bar{B} [u_{\mu},[u^{\mu},B]]  \rangle\no \\
&&
+ d_3 \langle \bar{B} u_{\mu} \rangle  \langle u^{\mu} B  \rangle
+ d_4 \langle \bar{B} B \rangle  \langle u^{\mu}  u_{\mu} \rangle .
\end{eqnarray}
Explicit chiral symmetry breaking is induced via the quark mass matrix
${\cal M} = \mbox{diag}(m_u, m_d,m_s)$
which enters in the combination $\chi_+ = 2 B_0 (u^\dagger {\cal M}
u^\dagger + u {\cal M} u  )$
with $B_0 = - \langle  0 | \bar{q} q | 0\rangle/ f^2$ representing the order
parameter of spontaneously broken chiral symmetry.

In the present work the numerical values for the couplings $b_i$ and $d_i$ have been constrained
as in the recent coupled channel analysis of ref.\cite{BMW} which includes $\eta$ photoproduction on nucleons as a high quality data set. We shall allow for small variations 
around the central values obtained in that work, for the following reason:
in the approach of \cite{BMW} only the contact interactions and the direct 
Born term for meson-baryon scattering were taken into account whereas, in addition, the
crossed Born term is included in the present analysis. We can therefore expect small changes in the numerical determination of the coupling constants from a fit to low-energy hadronic data.

In the current investigation we employ a relativistic chiral unitary approach 
to the strong $\bar{K} N$ interaction based on coupled channels which accounts for the important
contributions of the nearby $\Lambda(1405)$ resonance. 
By imposing constraints from unitarity we perform the
resummation of the amplitudes obtained from the
tree level amplitudes and the loop integrals.

The relativistic tree level amplitudes $V_{bj,ai}(s,\Omega;\sigma,\sigma')$ for
the meson-baryon scattering processes 
$B_{a}^{\sigma} \phi_i \rightarrow B_{b}^{\sigma'} \phi_j$ (with spin indices $\sigma$, $\sigma'$)
at leading chiral orders are obtained from both the contact interactions 
and the direct and crossed Born terms derived from the Lagrangian ${\cal L}$.
Since we are primarily concerned with a narrow center-of-mass energy region around
the $\bar{K} N$ threshold, it is sufficient to restrict ourselves to the 
$s$-wave (matrix) amplitude $V(s)$ which is given 
by
\begin{equation}
V(s) = \frac{1}{ 8 \pi} \sum_{\sigma =1}^2 \int d \Omega \; V(s,\Omega;\sigma, \sigma) ,
\end{equation}
where we have averaged over the spin $\sigma$ of the baryons and $s$ is
the invariant energy squared.

For each partial wave unitarity imposes a restriction on
the (inverse) $T$-matrix above the pertinent thresholds
\begin{equation} \label{unit}
\mbox{Im} T^{-1} = - \frac{|\mbox{\boldmath$q$}_{cm}|}{8 \pi \sqrt{s}} 
\end{equation}
with the three-momentum $\mbox{\boldmath$q$}_{cm}$ in the
center-of-mass frame of the channel under consideration.
Hence the imaginary part of $T^{-1}$ is 
given by the imaginary part
of the basic scalar loop integral $\tilde{G}$
above threshold,
\begin{equation}
\tilde{G}(q^2) = \int \frac{d^d l}{(2 \pi)^d} 
\frac{i}{[ (q-l)^2 - M_B^2 + i \epsilon]
   [ l^2 - m_\phi^2 + i \epsilon] } ~ ,
\end{equation}
where $M_B$ and $m_\phi$ are the physical masses of
the baryon and the meson, respectively.
For the finite part $G$ of $\tilde{G}$
one obtains, e.g., in dimensional regularization:
\begin{eqnarray}
G(q^2) &=& a({\mu}) + \frac{1}{32 \pi^2 q^2} \Bigg\{ q^2
\left[ \ln\Big(\frac{m_\phi^2 }{\mu^2}\Big) +
\ln\Big(\frac{M_B^2 }{\mu^2}\Big) -2 \right]\no \\
&& 
+ (m_\phi^2 - M_B^2)  \ln\left(\frac{m_\phi^2 }{M_B^2}\right)   
- 8 \sqrt{q^2}|\mbox{\boldmath$q$}_{cm}| \no \\
&& \qquad \times \mbox{artanh }
\left(\frac{2 \sqrt{q^2}|\mbox{\boldmath$q$}_{\scriptstyle{cm}}|}{
(m_\phi + M_B )^2 - q^2} \right) \Bigg\}~,
\end{eqnarray}
where $\mu$ is the regularization scale. 
The subtraction constant $a({\mu})$ cancels the scale dependence of the chiral logarithms
and simulates higher order contributions with
the value of $a({\mu})$ depending on the respective channel, cf.\cite{OM}. 

To the order we are working the inverse of the $T$ matrix is written as
\begin{equation} \label{invers}
T^{-1} = V^{-1} + G
\end{equation}
which yields  
\begin{equation} \label{V}
T = [1 + V \cdot G]^{-1} \; V \, .
\end{equation}
Eq. (\ref{V}) is understood as a matrix equation in each partial wave. 
The diagonal matrix $G$ collects the loop integrals in each channel.
This amounts to a summation of a bubble chain to all orders in the $s$-channel,
equivalent to solving a Bethe-Salpeter equation with $V$ as driving term.

We perform a global $\chi^2$ fit to a large amount of data, including
$K^- p$ scattering into coupled 
$S = -1$
channels, the threshold branching 
ratios of $K^- p$ into  $\pi \Sigma$ and $\pi^0\Lambda$ channels, the $\pi \Sigma$ mass spectrum, and the shift and width of kaonic hydrogen
recently measured at DEAR \cite{DEAR}.
The resulting values of the subtraction constants $a({\mu})$ at $\mu=1$ GeV are
$a_{\bar{K} N}({\mu}) = 0.95 \times 10^{-3}$, 
$a_{\pi \Lambda}({\mu}) = -0.59 \times 10^{-3}$, 
$a_{\pi \Sigma}({\mu}) = 1.80 \times 10^{-3}$, 
$a_{\eta \Lambda}({\mu}) = 2.92 \times 10^{-3}$, 
$a_{\eta \Sigma}({\mu}) = 0.98 \times 10^{-3}$, 
$a_{K \Xi}({\mu}) = 2.90 \times 10^{-3}$.
For the couplings $b_i, d_i$ we find (in units of GeV$^{-1}$)
$b_0 =-0.362 $, $b_D = 0.002 $, $b_F =-0.128 $,
and $d_1 =-0.11 $, $d_2 =0.05 $, $d_3 =0.31 $, $d_4 =-0.32 $.
%in units of GeV$^{-1}$.
%
The decay constant in the chiral limit, $f$, is varied between
the physical values of the pion decay constant $F_\pi = 92.4$~MeV, and the kaon
decay constant, $F_K= 112.7$~MeV,
since both pions and kaons are involved in the coupled channels. 
The present fit yields $f = 103.1$~MeV.

The Coulomb interaction has been shown to yield  significant 
contributions to the elastic $K^- p$ scattering amplitude up to kaon laboratory momenta of
100-150 MeV/$c$ \cite{JackDal}. Close to $K^- p$ threshold the electromagnetic meson-baryon
interactions are thus important and should not be neglected 
as in previous coupled channel calculations \cite{KSW2, OR, OM, LK, JOORM}.
We account for these corrections by adding the quantum
mechanical Coulomb scattering amplitude to the strong elastic $K^- p$ amplitude,
$f^{str}_{K^- p \to K^- p} = 1 / (8 \pi \sqrt{s})  T_{K^- p \to K^- p}^{str}$.
The total elastic cross section is then obtained by performing the integration over the
center-of-mass scattering angle. Since this integral is infrared divergent in the presence of the
Coulomb amplitude, a 
cutoff at extreme forward scattering angles must be introduced 
which we choose to be 
$\cos{\theta_{cm}} = 0.966$\,---\,the value employed in the data analyses of
refs.\,\cite{Hum, Sak}. The more detailed calculation will be 
presented in forthcoming work \cite{BNW}.

\begin{figure}[!h]
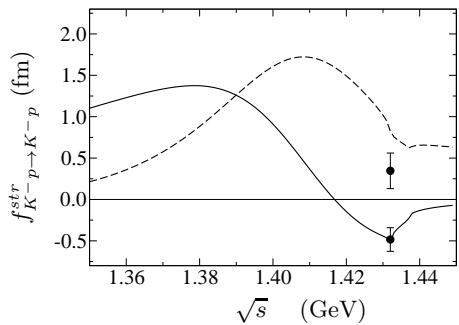

\centering
\begin{overpic}[width=0.3\textwidth,clip]{bestKmp.eps}
  \put(-11,15){\rotatebox{90}{{\scalebox{1.0}{$f^{str}_{K^- p \to K^- p}$ (fm)}}}}
  \put(45,-8){\scalebox{1.0}{$\sqrt{s}$ \quad (GeV)}}
\end{overpic}
\vspace{1.5ex}
\caption{Real (solid) and the imaginary part (dashed) of 
         the strong $K^- p \to K^- p$ amplitude, $f^{str}_{K^- p \to K^- p}$, as defined in the text. 
         The data points represent the real and imaginary parts of the $K^- p$ scattering length,
         derived from the DEAR experiment \cite{DEAR} with inclusion 
         of isospin breaking corrections according to ref.\,\cite{MRR}.}
\label{UbestKmp}
\end{figure}

\begin{figure}[!h]
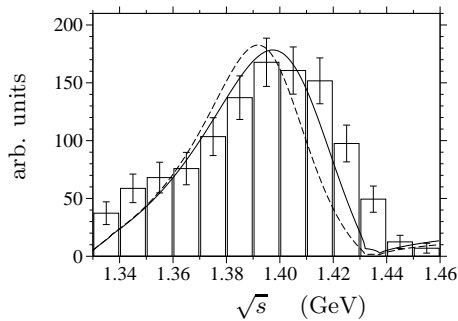

\centering
\begin{overpic}[width=0.3\textwidth,clip]{bestPiSigma.eps}
  \put(-11,25){\rotatebox{90}{{\scalebox{1.0}{arb. units}}}}
  \put(45,-8){\scalebox{1.0}{$\sqrt{s}$ \quad (GeV)}}
\end{overpic}
\vspace{1.5ex}
\caption{The $\pi \Sigma$ mass spectrum in the isospin $I = 0$ channel. The solid curve is 
         obtained from the overall $\chi^2$ fit to all available data. The dashed curve is found with the			additional constraint of remaining within the error margins of the DEAR data. The experimental
         histograms are taken from \cite{Hem}. The statistical errors have been supplemented following 		\cite{DD}.}
\label{UbestPiSigma}
\end{figure}

\begin{figure}[!h]
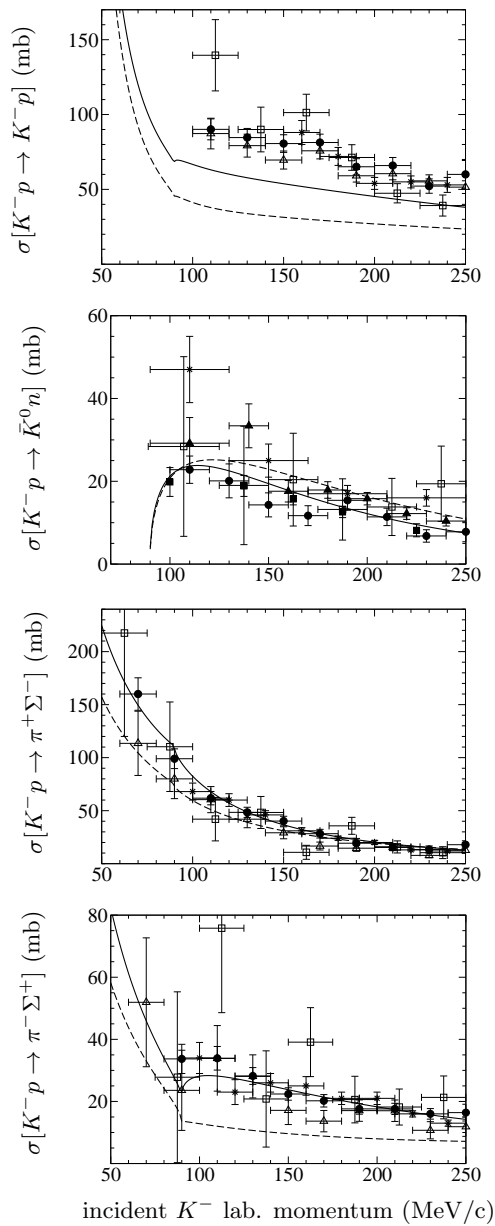

\centering
\begin{tabular}{r}
\begin{overpic}[height=0.155\textheight,clip]{bestCSKp.eps}
  \put(-14,7){\rotatebox{90}{{\scalebox{1.0}{$\sigma [K^- p \to K^- p]$ (mb)}}}}
\end{overpic} \\[0.25cm]
\begin{overpic}[height=0.155\textheight,clip]{bestCSKn.eps}
  \put(-17,8){\rotatebox{90}{{\scalebox{1.0}{$\sigma [K^- p \to \bar{K}^0 n]$ (mb)}}}}
\end{overpic} \\[0.25cm]
\begin{overpic}[height=0.155\textheight,clip]{bestCSPiSigm.eps}
  \put(-11,7){\rotatebox{90}{{\scalebox{1.0}{$\sigma [K^- p \to \pi^+ \Sigma^-]$ (mb)}}}}
\end{overpic} \\[0.25cm]
\begin{overpic}[height=0.155\textheight,clip]{bestCSPiSigp.eps}
  \put(-17,7){\rotatebox{90}{{\scalebox{1.0}{$\sigma [K^- p \to \pi^- \Sigma^+]$ (mb)}}}}
  \put(-1,-9){\scalebox{1.0}{incident $K^-$ lab. momentum (MeV/c)}}
\end{overpic}
\end{tabular}
\vspace{1.5ex}
\caption{Cross sections of $K^- p$ scattering into various channels
         obtained from the overall $\chi^2$ fit to all available data (solid curve) and with the 
         additional constraint of remaining within the DEAR data (dashed). The data
         are taken from  \cite{Hum} (empty squares), \cite{Sak} (empty triangles), 
         \cite{Kim} (full circles), \cite{Kit} (full squares), 
         \cite{Eva} (full triangles), \cite{Cib} (stars).}
\label{UbestCS}
\end{figure}

The results of the fit can be summarized as follows.
The strong-interaction part of the $K^- p$ amplitude, $f^{str}_{K^- p \to K^- p}$,
is presented in Figure~\ref{UbestKmp}.
At threshold we
obtain the $K^- p$ strong-interaction scattering length
\begin{equation}
a_{K^- p} = ( -0.51 + 0.82 \,i ) \ \mathrm{fm}.
\end{equation}
The $\pi \Sigma$ mass spectrum
in the isospin $I=0$ channel
is shown in Fig.~\ref{UbestPiSigma},
while the total cross sections of $K^- p$ scattering to various channels 
is displayed in Fig.~\ref{UbestCS}.

Additional tight constraints are provided by the well-measured threshold
ratios of the $K^- p$ system for which we find:
\beqa
\gamma & = & \frac{\Gamma(K^- p \to \pi^+ \Sigma^-)}{\Gamma(K^- p \to \pi^- \Sigma^+)} = 2.35~, 
 \no \\
R_c & = & \frac{\Gamma(K^- p \to \pi^+ \Sigma^- , \ \pi^- \Sigma^+)}
               {\Gamma(K^- p \to \textrm{\small all inelastic channels})} = 0.653~, \no \\
R_n & = & \frac{\Gamma(K^- p \to \pi^0 \Lambda)}{\Gamma(K^- p \to \textrm{\small neutral states})}
 = 0.194~. 
\eeqa
The experimental values $\gamma = 2.36 \pm 0.04, R_c = 0.664 \pm 0.011, 
R_n = 0.189 \pm 0.015$ \cite{Now, Tov} 
are perfectly well reproduced by our approach. (We mention in passing that this fit does not support 
a pronounced two-pole structure in the region of the $\Lambda(1405)$ as advocated in ref.\,\cite{JOORM}.)

\begin{figure}
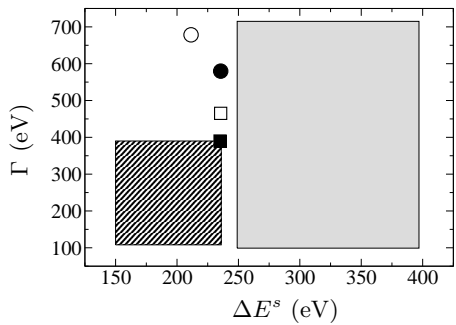

\centering
\begin{overpic}[width=0.3\textwidth]{bestDEAR.eps}
  \put(-10,27){\rotatebox{90}{\scalebox{1.0}{$\Gamma$ (eV)}}}
  \put(45,-8){\scalebox{1.0}{$\Delta E^s$ (eV)}}
\end{overpic}
\vspace{1.5ex}
\caption{Results for the strong interaction shift and width of kaonic hydrogen
         from our approach, both by using the Deser-Trueman formula \cite{DT} (empty circle)
         and by including isospin breaking corrections \cite{MRR} (full circle).
         The DEAR data are represented by the shaded
         box \cite{DEAR}, the KEK data by the light gray box \cite{Ito}.
         The fit restricted to the DEAR data is represented by the small full rectangle 
         (empty rectangle without isospin breaking corrections).}
\label{UbestDEAR}
\end{figure}

It turns out, however,  that these results cannot be brought to simultaneous satisfactory
agreement with the elastic $K^- p$ total cross section, and with the strong
interaction shift and width in kaonic hydrogen measured at DEAR \cite{DEAR}. We find $\Delta E = 236$ eV 
and $\Gamma = 580$ eV (with inclusion of isospin breaking corrections following \cite{MRR}), 
see Fig.~\ref{UbestDEAR}.
In comparison with previous coupled-channels calculations, the situation is ameliorated 
by including electromagnetic corrections to $K^- p$ scattering which are important close to threshold.
Nevertheless, inclusion of the Coulomb interaction cannot account for the apparent gap between
the DEAR result and the bulk of the existing elastic $K^- p$ scattering data (the latter are, admittedly, 
of low precision). While there is consistency with the new value for the energy shift in kaonic hydrogen, 
it is now difficult to accomodate the scattering data with the much improved accuracy of the measured 
width \cite{DEAR}. 
(Note that the radiative decays of $K^- p$ into $\Lambda \gamma$ and $\Sigma^0 \gamma$ are expected
to contribute less than 1\% to the decay width of kaonic hydrogen
and can be safely omitted \cite{Iva}.)

The results of our calculations represent an ``optimal'' compromise between the various existing data sets.
If, on the other hand, one imposes the constraint of remaining strictly within the error band of the 
DEAR data,
the fit yields $\Delta E = 235$ eV, $\Gamma = 390$ eV which corresponds to
a strong-interaction scattering length $a_{K^- p} = ( -0.57 + 0.56 \,i ) $ fm. With this constraint imposed, 
we obtain $\gamma = 2.38, R_c = 0.631$, $R_n = 0.176$ and 
a shifted $\pi \Sigma$ mass spectrum (dashed curve in \ Fig.~\ref{UbestPiSigma}), while the calculated $K^- p$ scattering cross sections move to the dashed curves in Fig.~\ref{UbestCS}.

We have also performed fits omitting the DEAR results. Our calculations are then in good agreement with all scattering data including the elastic $K^- p$ channel. The fits are also
within the (larger) error bars of the previous KEK measurement \cite{Ito}. We have furthermore convinced ourselves that we obtain qualitatively similar results by applying several variants of the approach presented here: first by using only the 
Weinberg-Tomozawa part of the
driving term $V$, then by adding subsequently the higher
order contact interactions $b_i, d_i$ and the direct Born term. These studies will be discussed in detail in a forthcoming report \cite{BNW}. 

In conclusion, the present updated analysis of low-energy $K^-$-- proton interactions, 
combining the next-to-leading order chiral SU(3) effective Lagrangian with an improved 
coupled-channels approach, emphasizes the importance of the constraints set by the new 
accurate kaonic hydrogen data from the DEAR experiment. At the same time this analysis 
points to questions of consistency with previously measured sets of $K^- p$ scattering 
data. Developments aiming for a precision at the level
of a few electron volts in the shift and width of kaonic hydrogen, foreseen at DA$\Phi$NE 
in the near future, will further clarify the situation. \\

We thank C.~Guaraldo, D.~Jido, N.~Kaiser, P.~Kienle and A.~Rusetsky
for useful discussions. This work was supported in part by DFG and BMBF.

%%%%%%%%%%%%%%%%%%%%%%%%%%%%%%%%%%%%%%%%%%%%%%%%%%%%%%%%%%%%%%%%%%%%%%%%%%%%%%

\end{document}